\documentclass[preprint]{ptephy_v1}%%%%%% to generate preprint number with ptep logo

\preprintnumber{XXXX-XXXX} %%% %%% Insert preprint number here
\usepackage{hyperref}
\usepackage{subfig}
\usepackage{caption}
\usepackage{ulem}
\begin{document}

\title{Performance Calculation of Pulse Shape Discrimination Based on Photoelectron Quantity}

%%%% To generate auto affiliation numbers please use \author{}\affil{} command

\author{S. B. Hong}
\author{J. S. Park*}
\affil{Kyungpook National University \email{jungsicpark@knu.ac.kr}}

%\author{J. S. Park}
%\affil{Insert second author address here}

%\author{Insert third author name here}
%\author[3]{Insert fourth author name here} %%% Use optional bracket [3] to change the respective address
%\affil{Insert third author address here}

%\author{Insert last author name here\thanks{These authors contributed equally to this work}}
%\affil{Insert last author address here}

%%% To include the collaborator name... Please use the command "\collaborator"
%%% For example: \collaborator{ATLAS Collaboration}

\begin{abstract}%

Pulse Shape Discrimination (PSD) is a widely used technique in many experimental analysis. In this study, we specifically aimed to assess the effectiveness of PSD in accurately measuring decay time. We measured the decay times of a 0.1 wt\% Gd-loaded liquid scintillator (Gd-LS) with 5 vol\% Ultimagold-F added when irradiated with neutrons and gamma rays, which were emitted from a Cf-252 radiation source, using a two-exponential decay model. We distinguished between gamma-like events and neutron-like events using the time-of-flight difference. Based on the measured decay times, we developed a simulation to model the waveforms. In the simulation, we adjusted the number of photoelectrons (NPE) and generated waveforms for NPE ranging from 10 to 1000 photoelectrons. We investigated the pulse shape discrimination (PSD) performance as a function of NPE photoelectrons (PE) and determined that at least 49 PE is required for a neutron-like events rejection efficiency of 90\%, while keeping gamma-like events 97.8\%.

%Ver1-----------------------------------------------------------------------------------------------------------
%We developed a simulation to model the waveforms of neutron and gamma events for a 0.1 wt.\% Gd-LS with 5 vol.\% UltimaGold-F. The decay times were measured as 10 ns for the fast component and 53 ns for the slow component for gamma events, and 14 ns and 63 ns, respectively, for neutron events. The simulation predicted that to achieve 90\% performance using the Log-likelihood methodology, a minimum of 47.6 PE is required, while 95\% performance necessitates at least 86.7 PE.
%------------------------------------------------------------------------------------------------------------------
% We measured decay times of Gd-LS(gadolinium-loaded liquid scintillator). 
% - decay time is measured -> fitting result ( gamma & neutron)
% - A simulation that replicate the waveform has been made by using measured decay time
% -  simulation predicts ~~% performance ~~~, ~~~ ~~~ %

\end{abstract}

\subjectindex{xxxx, xxx}

\maketitle

%Introduction------------------------------------------------------------------------------------------------------------
\section{Introduction}
%This demo file is intended to serve as a ``starter file''
The pulse shape discrimination (PSD) method is widely utilized in data analysis to distringuish between various types of radiation, particularly in experiments using scintillator cocktails and crystals~\cite{cite:NEOS,cite:PROSPECT, cite:COSINE, cite:JSNS2}. Many of these experiments rely on data obtained from the inverse beta decay (IBD) process within metal-loaded Liquid Scintillator (LS). In this process, electron antineutrinos interact with free protons in the scintillator. The resulting positron and neutron then deposit energy into the medium through annihilation and capture, respectively. These energy depositions generate two distinct signals: a prompt signal and a delayed signal. The prompt signal is produced by the annihilation of the positron, while the delayed signal results from the capture of the neutron by metal nuclei. The delayed signal can be distinguished from environmental background noise by its characteristic energy. This enables the detection of both prompt and delayed signals when IBD occurs within the metal-loaded LS, with the two signals separated by a short time interval. This separation allows time-coincidence logic to distinguish genuine IBD events from accidental coincidences. However, fast neutrons pose a significant challenge, as they can produce signals that resemble IBD. When a fast neutron interacts with the scintillator, it undergoes two processes that result in similar false signals. First, the fast neutron scatters off a proton, generating a recoil proton that produces a false prompt signal. Subsequently, the fast neutron is captured by metal nuclei, leading to a false delayed signal. Since these false signals occur with time intervals comparable to real IBD events, it is difficult to distinguish genuine signals from background noise using time-coincidence logic alone. In such cases, the PSD method is crucial for distinguishing between real and false events. We measured the decay times of neutron and gamma events in a 0.1 wt\% Gd-LS which is produced based on~\cite{cite:Gd-LS} in addition to 5 vol\% UltimaGold-F~\cite{cite:UltimaGold_F}. Using a Cf-252 source, neutrons and gamma rays were introduced into the scintillator and the time-of-flight (ToF) difference was employed to distinguish between the two events. Based on the measured decay times, we developed a simulation to model the waveforms, and by varying the number of photoelectrons (NPE) in the simulation, we explored the performance of the PSD methodology. In Section 2, the measurement of decay time is discussed, including its required experimental setup, event selection logic, and data analysis process. Section 3 addresses the waveform simulation, introducing the factors considered in generating Monte Carlo (MC) simulation and the elements calibrated to match real-world conditions.

%\verb+ptephy_v1.cls+ v0.1

%\subsection{Insert B head here}
%Subsection text here.

%\subsubsection{Insert C head here}
%Subsubsection text here.

%----------------------------------------------------------------------------------------------------------------------------

%------------------------------------------------------------------------------------------------------------------------------
\section{Measurement of Decay time}

\subsection{Measurement setup}

\begin{figure}[htbp]
    \centering
    % first subfigure
    \subfloat[Schematic diagram of experimental setup.\label{fig:exp_setup_1}]{
        \includegraphics[width=0.35\textwidth]{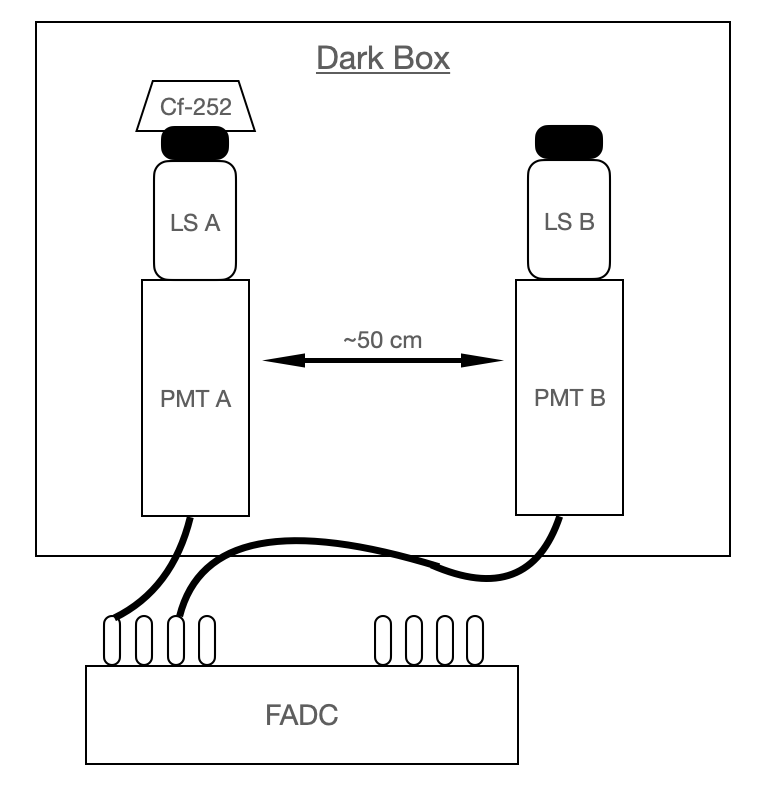}
    }
    \hfill
    % second subfigure
    \subfloat[Data acquisition system.\label{fig:exp_setup_2}]{
        \includegraphics[width=0.45\textwidth]{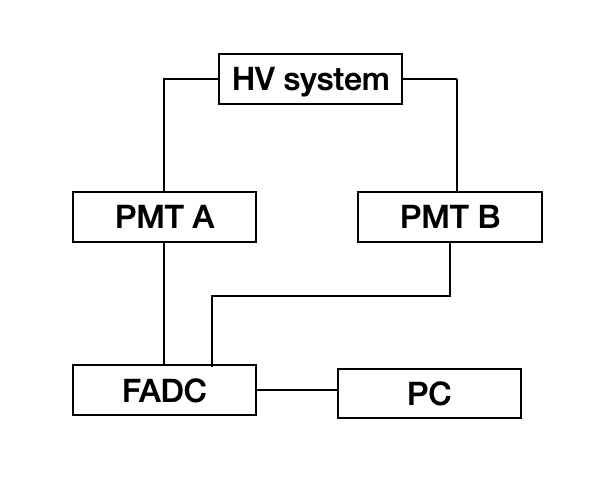}
    }
    
    \caption{(a) The interior of the dark box. LS B and the Cf-252 source are separated by approximately 50 cm. (b) Data acquisition process. A voltage of 1600 V is applied to each PMT, and the PMT A generates a trigger to store both PMT waveforms.}
    \label{fig:main_exp_setup}
\end{figure}

The experimental setup consists of two photomultiplier tubes (PMTs), two different liquid scintillators, a flash analog to digital converter (FADC), a high voltage (HV) supplier, a Cf-252 radiation source, and a dark box as shown in Fig.~\ref{fig:main_exp_setup} (a)(b). Two 2-inch Hamamatsu H7195 PMTs~\cite{cite:Hamamatsu_H7195} were used. A NOTICE NK-FADC with a 500 MHz sampling rate was employed, offering a 2.5V dynamic range with 12-bit resolution. A CAEN SY4527~\cite{cite:CAEN_SY4527} power supply unit equipped with a CAEN Mod. A1733N board was used to supply 1600 V to both PMTs.

Inside the dark box, two PMTs were installed at a distance of approximately 50 cm apart. Above each PMT, the LS was vertically attached; for the two PMTs, we refer to them in this work as "PMT A with LS A" and "PMT B with LS B". A Cf-252 radioactive source was positioned above LS A. The liquid scintillator used Linear Alkyl Benzene (LAB) as the solvent, with PPO (3 g/L) and bis-MSB (0.03 g/L) added. Additionally, LS B was loaded with 0.1 wt\% gadolinium and 5 vol\% UltimaGold-F. The waveform from PMT A was used as a trigger, allowing the waveforms from PMT A and PMT B to be recorded simultaneously.

The neutrons and gamma rays emitted from Cf-252 differ in velocity, with neutrons from Cf-252 having about 2 MeV~\cite{cite:Cf}, corresponding to a velocity of approximately 2 cm/ns. In contrast, the gamma rays have a velocity of about 30 cm/ns, independent of their energy. Therefore, by positioning LS B approximately 50 cm away from the Cf-252 radiation source, we can use the time-of-flight (ToF) difference to distinguish between gamma and neutron events. From this 50 cm distance, a ToF difference of approximately 25 ns is expected. To analyze these events, we calculate the ToF difference between PMT A and PMT B.

\subsection{Event selection}

\begin{figure}[htbp]
    \centering
    % first subfigure
    \subfloat[Example waveforms of PMT A and B.\label{fig:ToF_PMT_A_and_B}]{
        \includegraphics[width=0.45\textwidth]{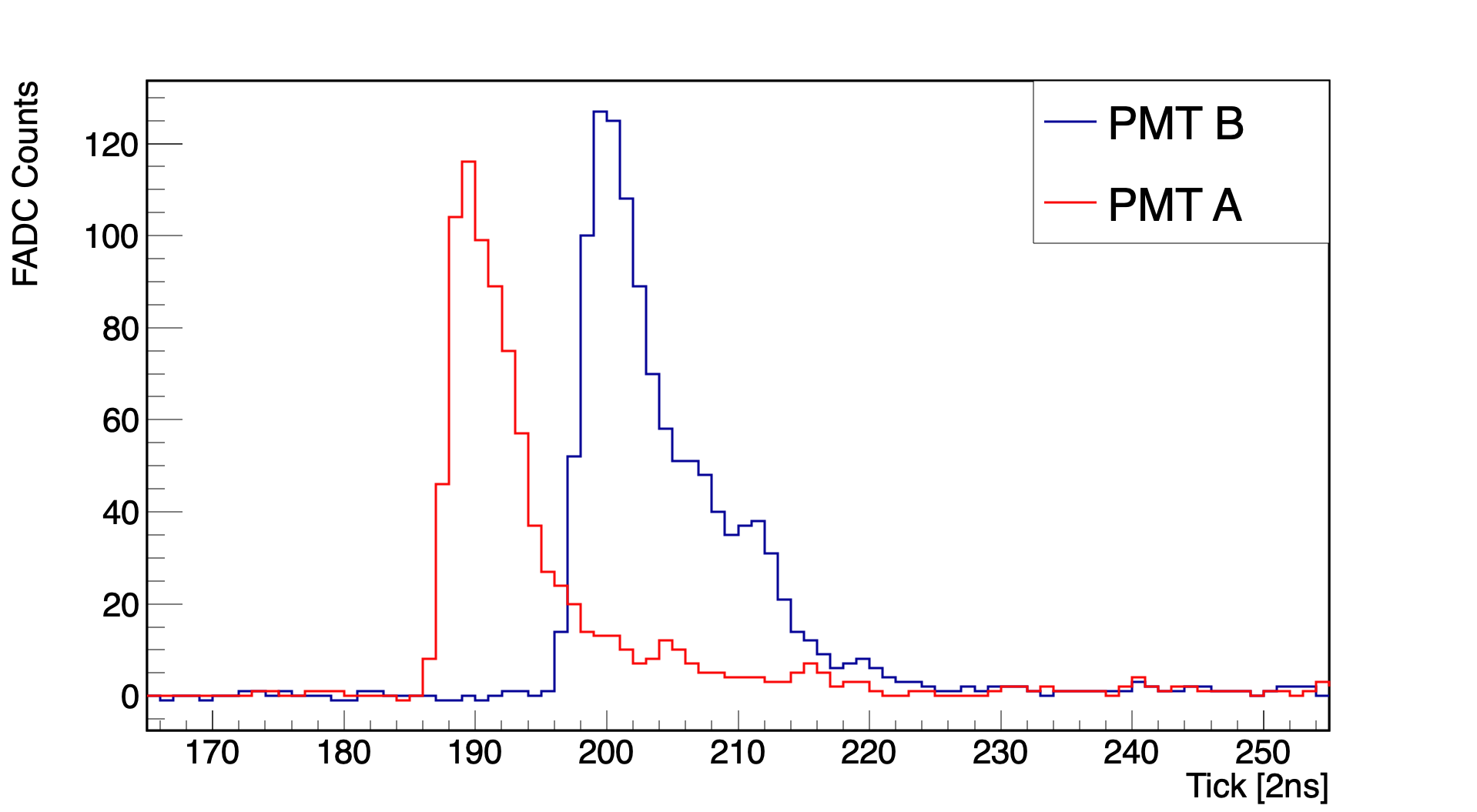}
    }
    \hfill
    % second subfigure
    \subfloat[Distribution of time-of-flight differences.\label{fig:graph1}]{
        \includegraphics[width=0.45\textwidth]{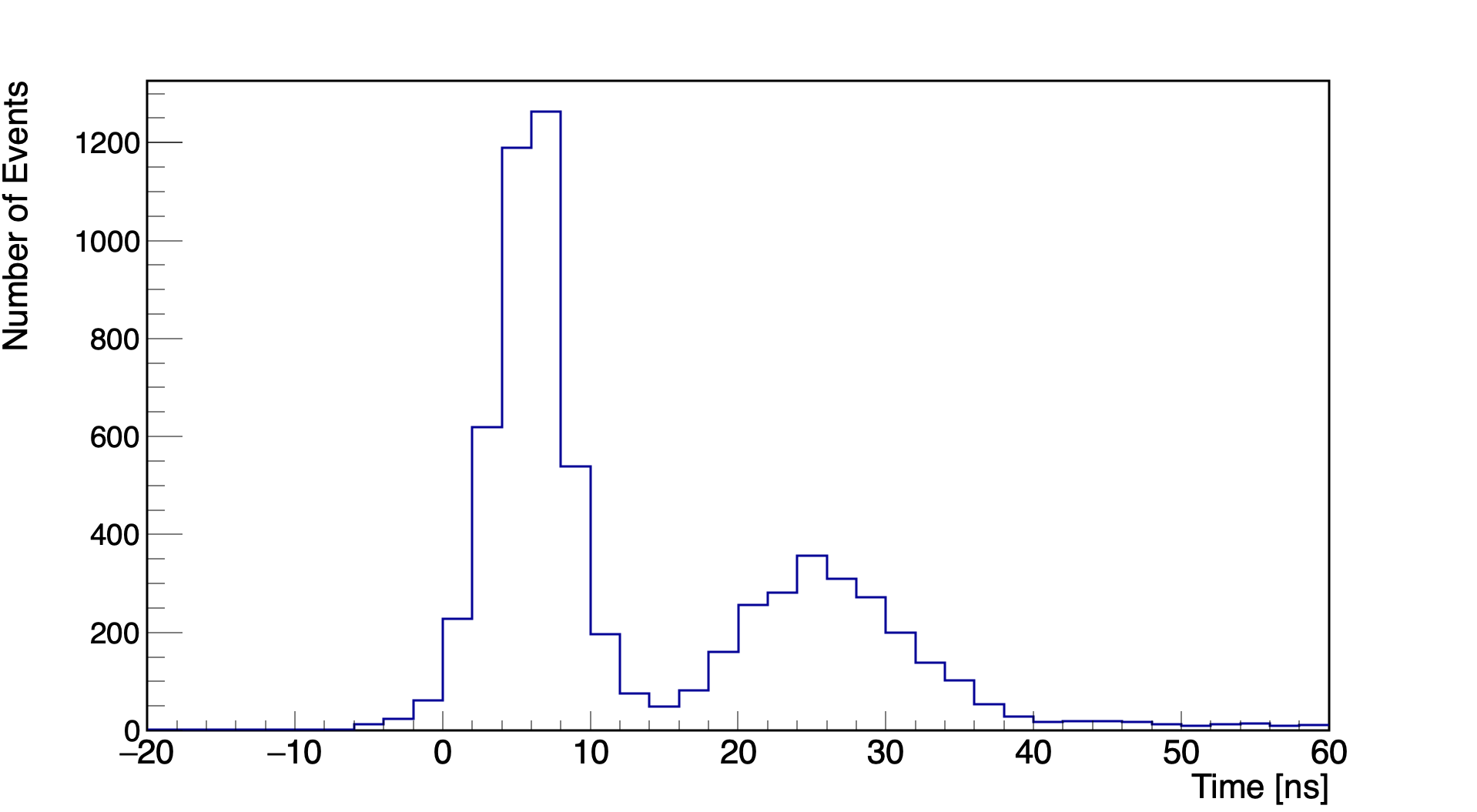}
    }
    \hfill
    % third subfigure
    \subfloat[Accumulated waveforms.\label{fig:graph2}]{
        \includegraphics[width=0.45\textwidth]{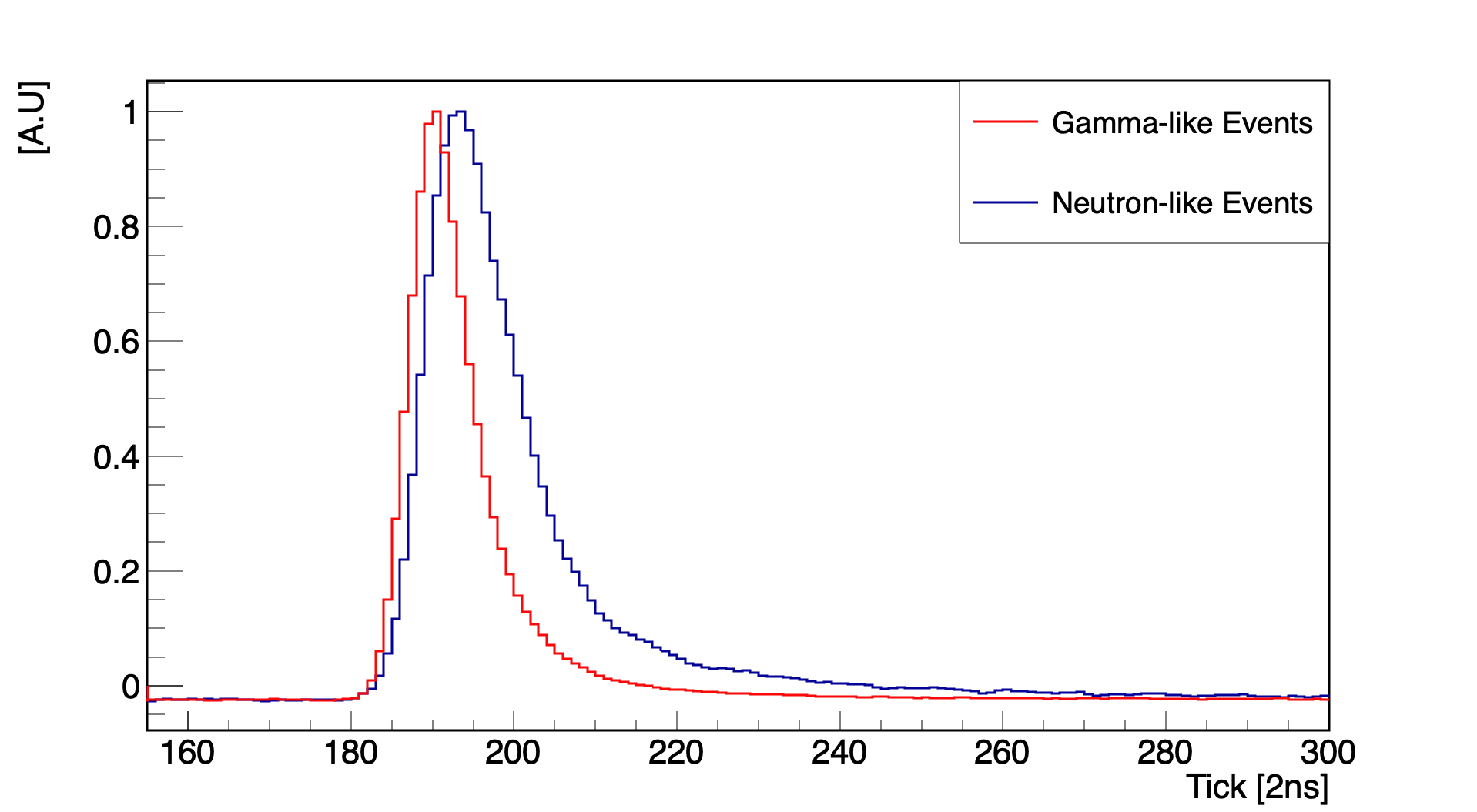}
    }
    
    \caption{(a) Case where a signal is also present in PMT B when triggered by PMT A. (b) Dsitribution of ToF differences. (c) Accumulated waveforms obtained from PMT A and PMT B.}
    \label{fig:main_graphs}
\end{figure}

From the experimental measurement results, we initially focused on events in which both PMT A and PMT B recorded waveforms with pulse heights exceeding 10 FADC counts. The ToF defined as the time difference between the peak of the waveform in PMT A and the peak in PMT B, was calculated, with its distribution illustrated in Fig. 2(b). As anticipated, a notable number of neutron-like events was observed at a ToF of approximately 25 ns. We further refined our analysis by selecting events with ToF values ranging from -10 ns to 40 ns. Within this range, events with a ToF less than 12 ns were classified as gamma-like, while those with a ToF greater than 12 ns were identified as neutron-like. The accumulated waveforms corresponding to these classifications are presented in Fig. 2(c).

\subsection{Data analysis}

\begin{equation}
	A_{1} \cdot \exp\left(-\frac{t}{\tau_{f}}\right) + A_{2} \cdot \exp\left(-\frac{t}{\tau_{s}}\right)
	\label{eq:decay}
\end{equation}

\begin{table}[!h]
	\caption{Decaytimes}%%%Table caption goes here
	\label{table_example}
	\centering
	\begin{tabular}{|c|c||c|c|c|}%%%The number of columns has to be defined here

	\hline
	Decay type & Material & Fast Component & Slow Component \\ %	%%% Table body
	\hline
	gamma & LS B & 10 & 53\\
	\hline
	neutron & LS B & 11 & 63\\
	\hline
	\end{tabular}
	\label{tab:fit_result}
\end{table}%%%End of the table

\begin{figure}[htbp]
    \centering
    % first subfigure
    \includegraphics[width=0.5\textwidth]{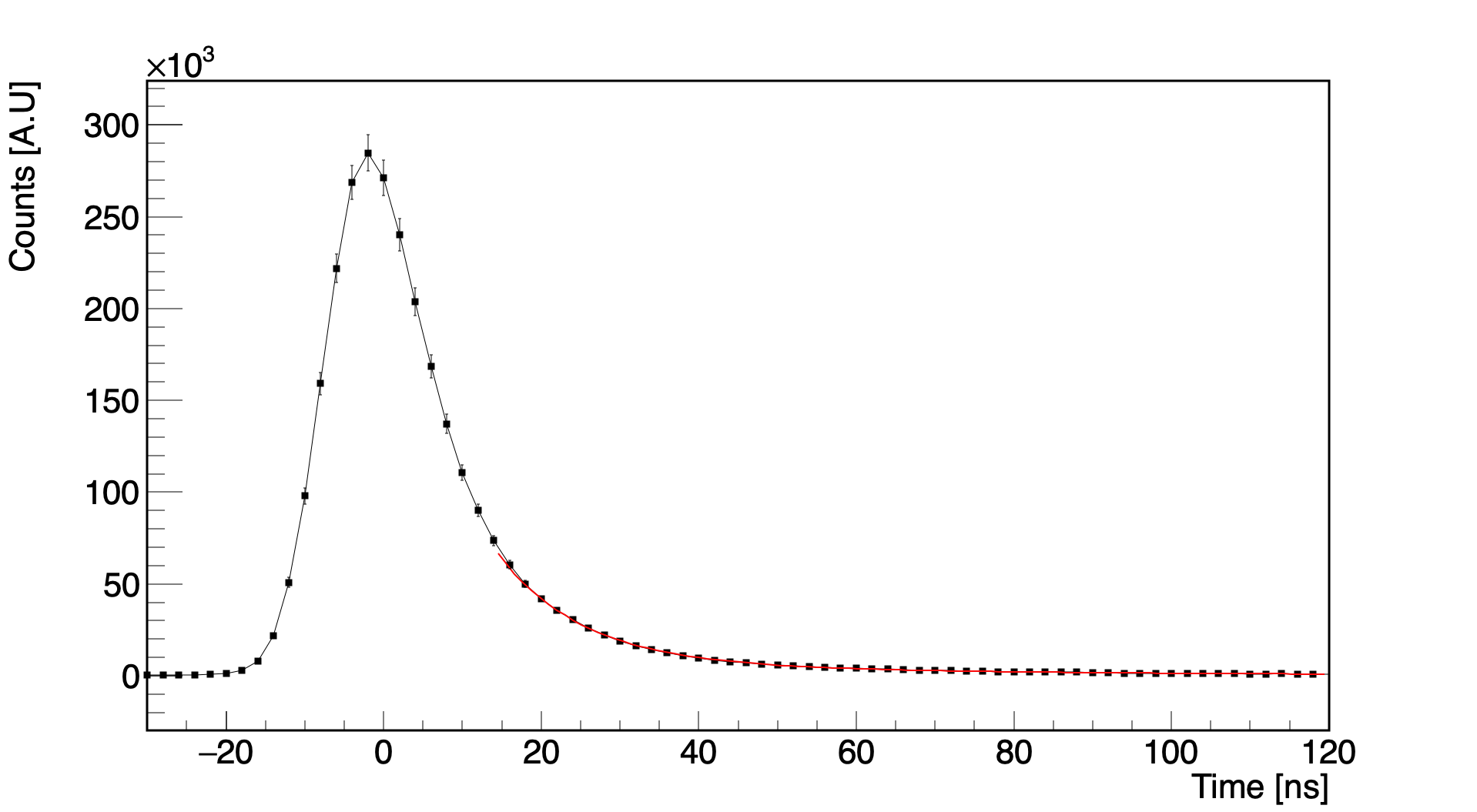}
    \hfill
    \caption{Graph fitting. The fitting interval ranges from 16 ns (8 ticks) to 180 ns (90 ticks) away from the peak of the waveform.}
    \label{fig:exp_Fit_grp}
\end{figure}

We fitted the accumulated waveform shown in Fig.~\ref{fig:graph2} using Equation 1. In this Equation, \( A_{1} \) and \( A_{2} \) represent the amplitudes of the two exponential decay components, corresponding to different decay processes occurring in the Cf-252. The parameters \( \tau_{f} \) and \( \tau_{s} \) are the characteristic decay times associated with these processes. Specifically, \( \tau_{f} \) denotes the fast component and \( \tau_{s} \) signifies the slow component. The results are presented in Table~\ref{tab:fit_result}. The fitting was performed from 8 ticks to 90 ticks away from the peak using the Minuit MIGRAD minimizer in CERN ROOT~\cite{cite:CERN_ROOT}.

%------------------------------------------------------------------------------------------------------------------------------

%------------------------------------------------------------------------------------------------------------------------------
\section{Pulse Shape Discrimination}
\subsection{Monte-Carlo Simulation}

\begin{figure}[htbp]
    \centering
    % first subfigure
    \subfloat[Waveform of a single photoelectron.\label{fig:SPE_waveform}]{
        \includegraphics[width=0.5\textwidth]{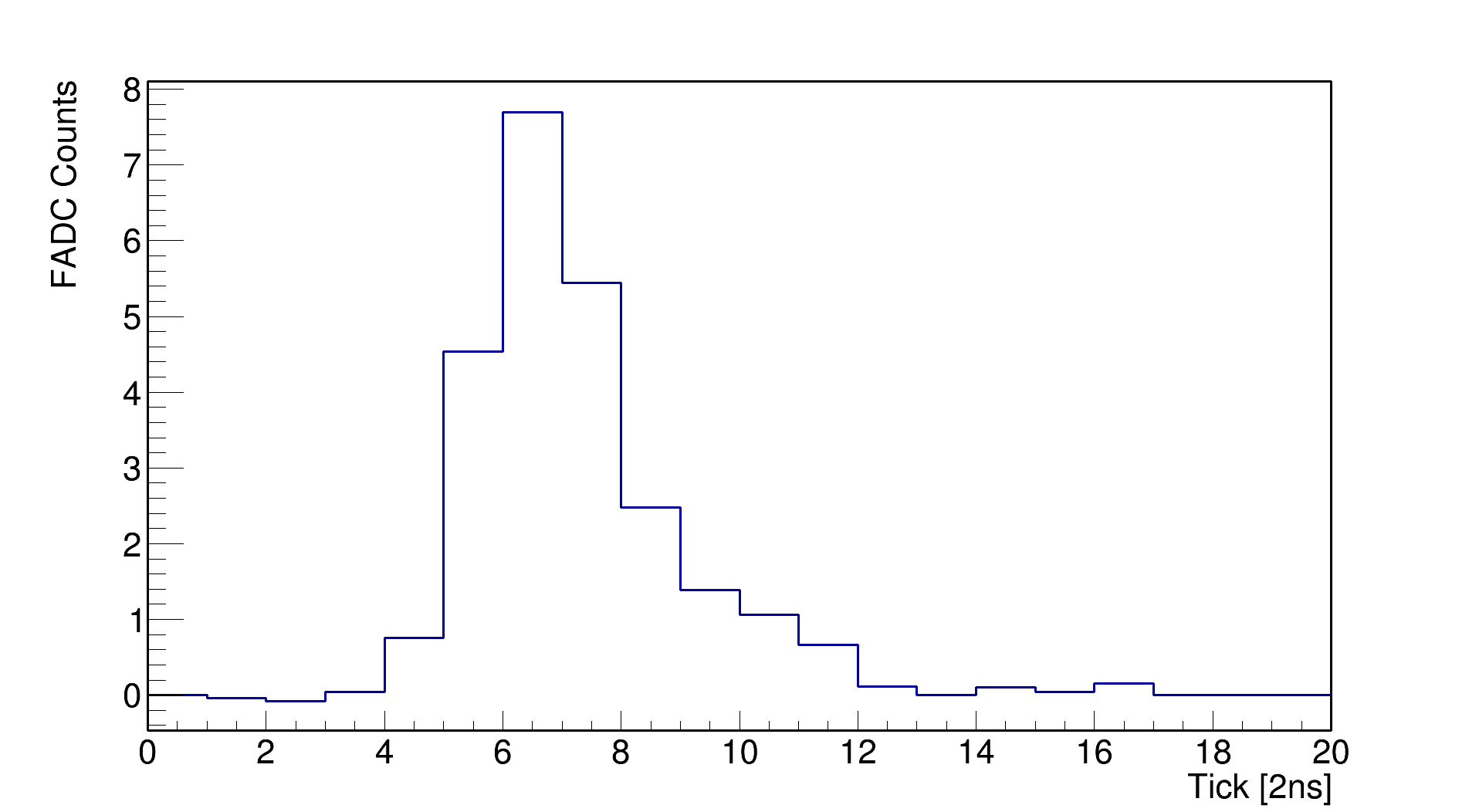}
    }
    \hfill
    % second subfigure
    \subfloat[Reproduced MC waveform based on NPE.\label{fig:NPE_charge_reproduce}]{
        \includegraphics[width=0.5\textwidth]{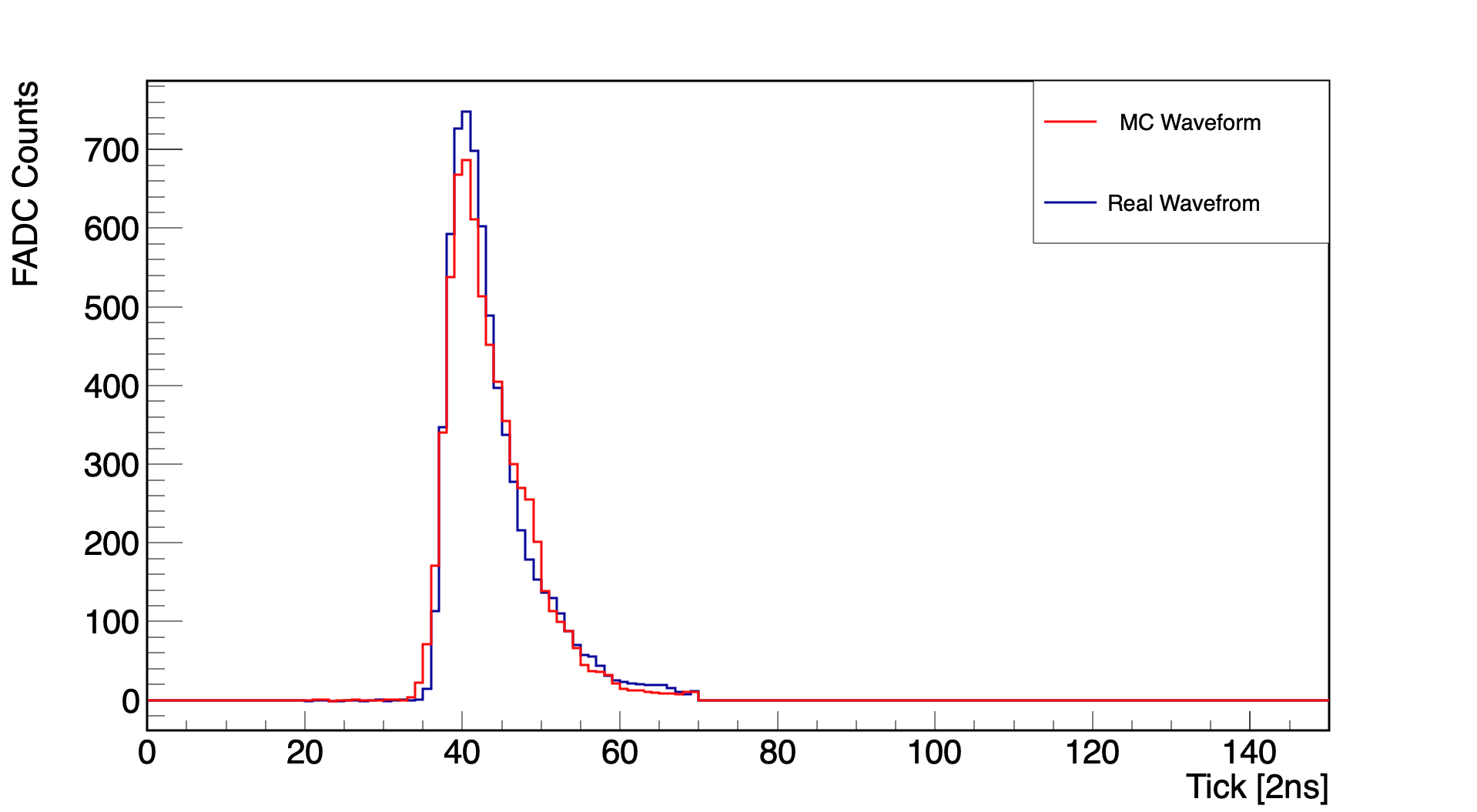}
    }
     \subfloat[Comparison of waveform between the MC and the data.\label{fig:stack_waveform_MC}]{
        \includegraphics[width=0.5\textwidth]{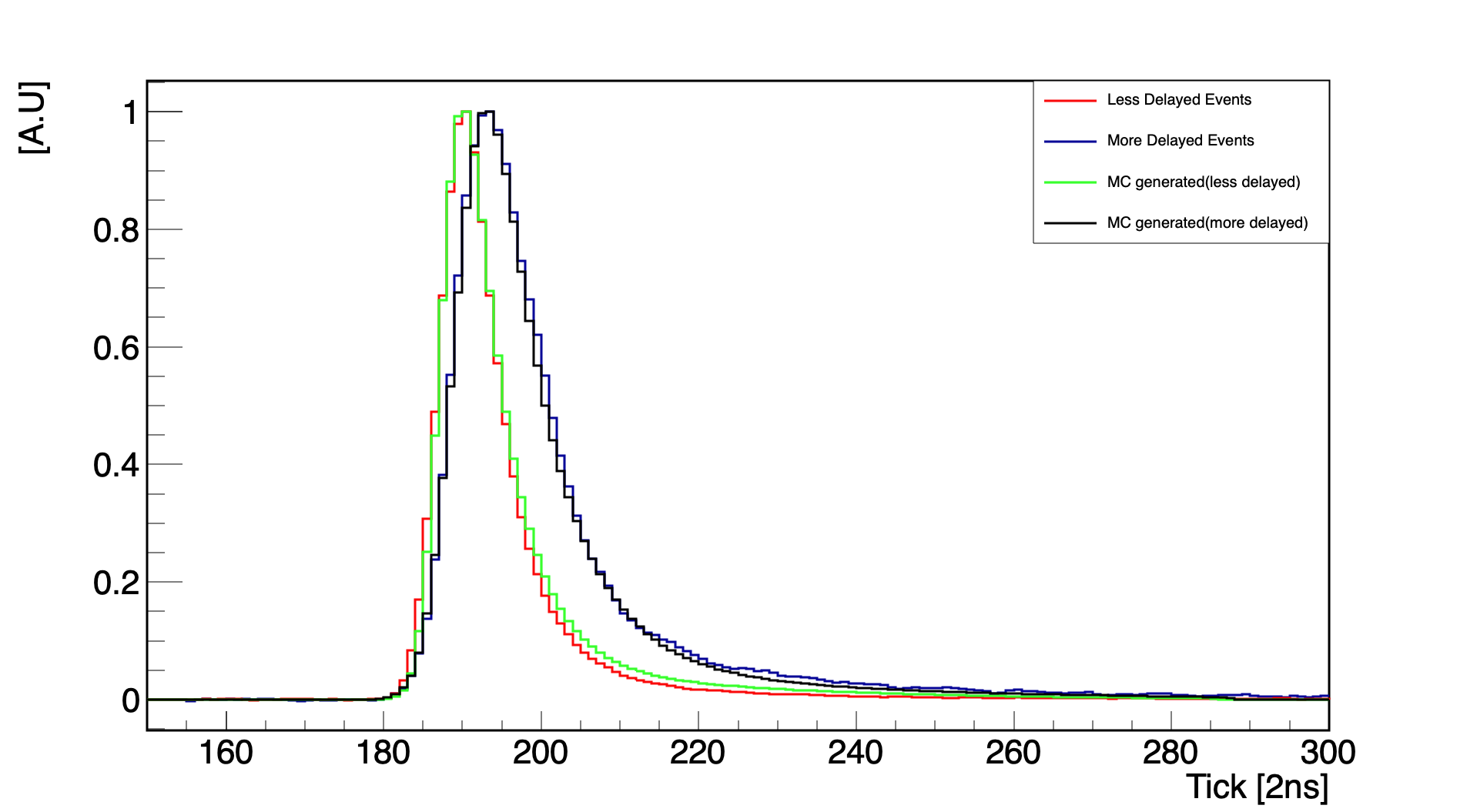}
    }
    
    \caption{(a) The waveform of the SPE. The SPE waveform was obtained through PMT dark-hit collection. (b) The waveform generated by the MC. (c) Comparison between the simulation and the data.}
    \label{fig:main_MC_generation}
\end{figure}

Our developed MC simulation models the waveforms of the PMT digitized by the FADC utilizing transit time spread (TTS), pedestal fluctuation, single photoelectron (SPE) effect, and the number of photoelectrons (NPE). We can simulate the desired waveforms by varying the NPE and decay time. Events were generated to replicate the decay times obtained in Section 2.2, with 10,000 events created for each NPE value ranging from 50 PE to 1,000 PE with a step size of 10 PE.

We accumulated approximately 5,000 PMT dark-hit waveforms to obtain the SPE waveform, as shown in Fig.~\ref{fig:SPE_waveform}. Events with a single peak in the dark-hit waveform with a pulse height between 5 and 15 FADC counts were selected. The simulation was then designed to accumulate PE based on the shape of the SPE waveform. Using data acquired from PMT B, we selected an arbitrary event and calculated the NPE based on the SPE charge. When the event was reproduced in the simulation according to the calculated NPE, the simulation data agreed well with the data, as illustrated in Fig.~\ref{fig:NPE_charge_reproduce}. As shown in Fig.~\ref{fig:stack_waveform_MC}, the MC can effectively reproduce both gamma-like and neutron-like events.

\subsection{Variance in Photoelectron Quantity}

In cases where multiple photoelectrons are generated, the likelihood function can be defined as follows:
\begin{equation}
L(\gamma) = \prod_{i=1}^{n} P_\gamma(t_i)
\end{equation}

The \( L(\gamma) \) denotes the likelihood function and \( P_\gamma \) represents the probability density function (PDF) of the gamma waveform according to the elapsed time. The function \( P_\gamma (t_i) \) accounts for the probability of the observed data point \( t_i \).

By taking the logarithm of both sides, we can transform the above likelihood function into a more accessible form:

\begin{equation}
\log(L(\gamma)) = \log\left( \prod_{i=1}^{n} P_\gamma(t_i) \right) = \sum_{i=1}^{n} \log[P_\gamma(t_i)]
\end{equation}

When the waveform is received via FADC, it is stored in the form of a histogram, thus the likelihood function must be modified as follows:

\begin{equation}
\log[L(\gamma)] = \sum_{i=1}^{\text{number of bins}} N_i \log[P_\gamma(t_i)]
\end{equation}

The log-likelihood function is now transformed into the summation of the likelihood function multiplied by corresponding counts of bin(\( N_i \)).

By applying the likelihood functions of the gamma and neutron waveform PDFs to the waveform of a single event, and then calculating the difference between these values, it is possible to determine the type of particle that caused the event.

\begin{equation}
\log[L(\gamma)] - \log[L(n)]
\end{equation}

Therefore, if the log-likelihood difference in the above equation is positive, it can be concluded that the event is a gamma event, with \( n \) denoting the neutron.

\begin{table}[!h]
\caption{PSD performance according to NPE}%%%Table caption goes here
\label{table_example}
\centering
\begin{tabular}{|c|c||c|}%%%The number of columns has to be defined here

\hline
desired PSD performance [\%] & Retainable gamma-like events [\%] & required NPE\\ %%%% Table body
\hline
90 &  97.8 & 49\\
\hline
95 & 99.4 & 79\\
\hline
99 & 99.9 &150\\
\hline
\end{tabular}
\end{table}%%%End of the table

\begin{figure}[htbp]
    \centering
        % first subfigure
    \subfloat[50PE.\label{fig:50PE}]{
        \includegraphics[width=0.5\textwidth]{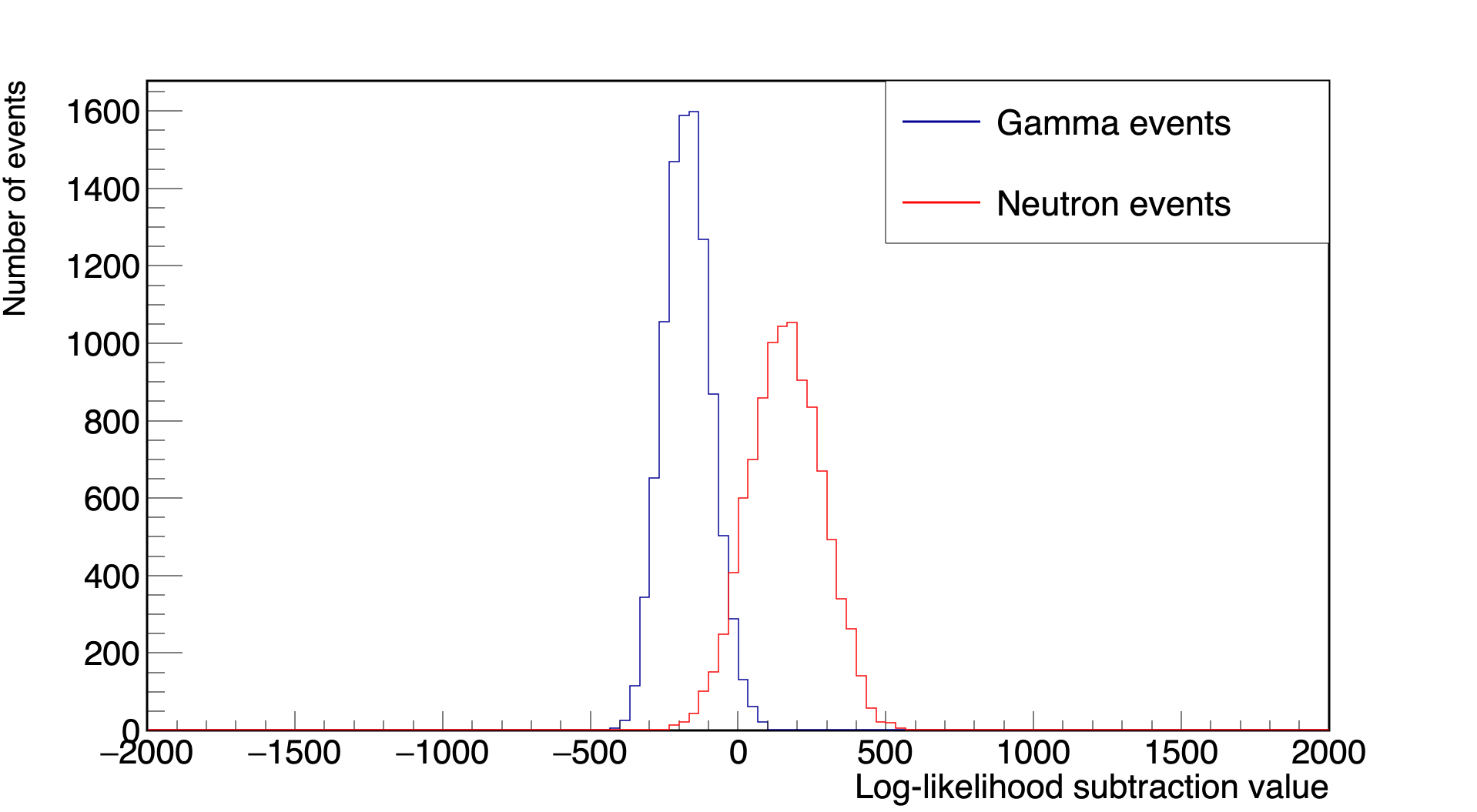}
    }
    \hfill
    % second subfigure
    \subfloat[80PE.\label{fig:80PE}]{
        \includegraphics[width=0.5\textwidth]{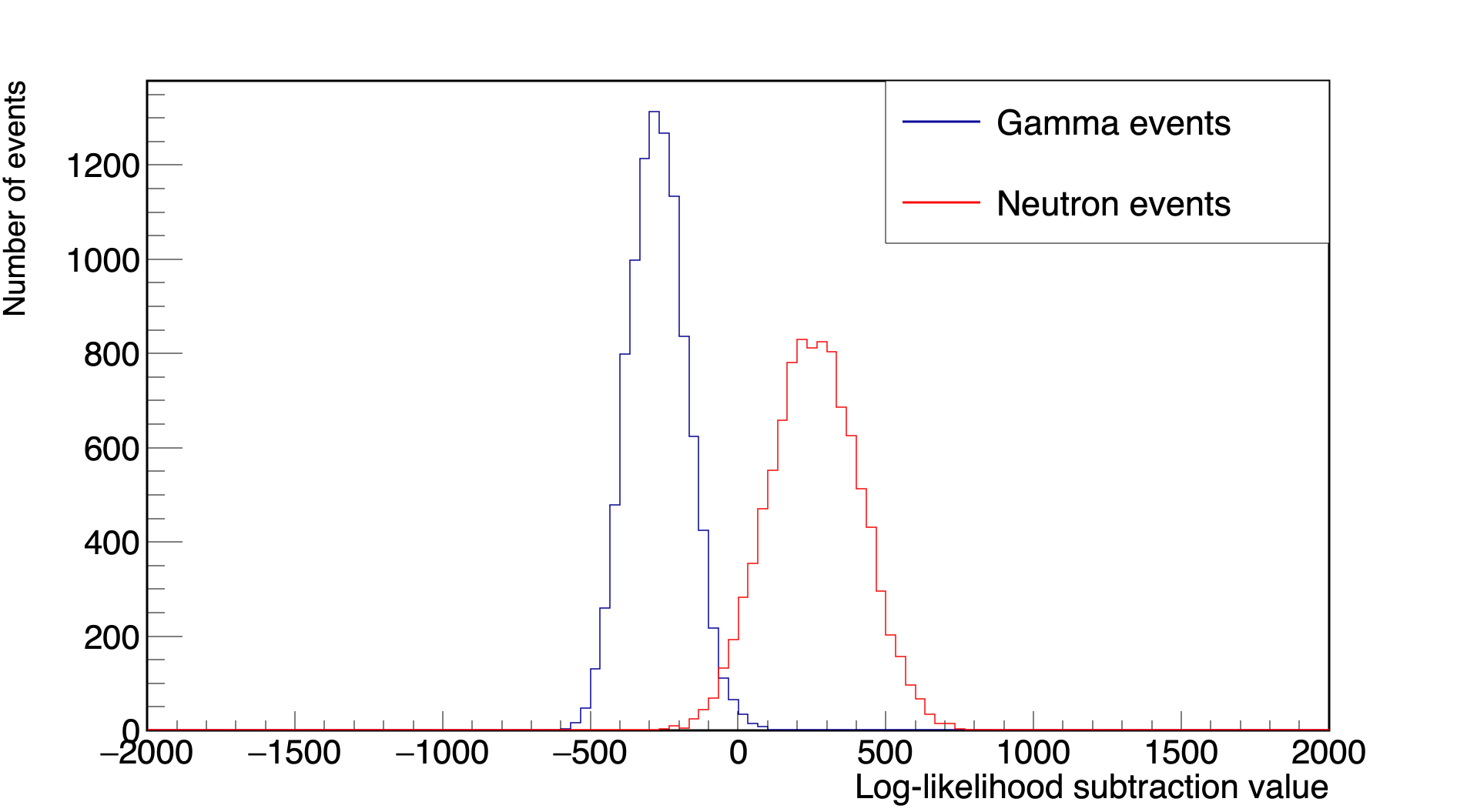}
    }
    % third subfigure
    \subfloat[150PE.\label{fig:150PE}]{
        \includegraphics[width=0.5\textwidth]{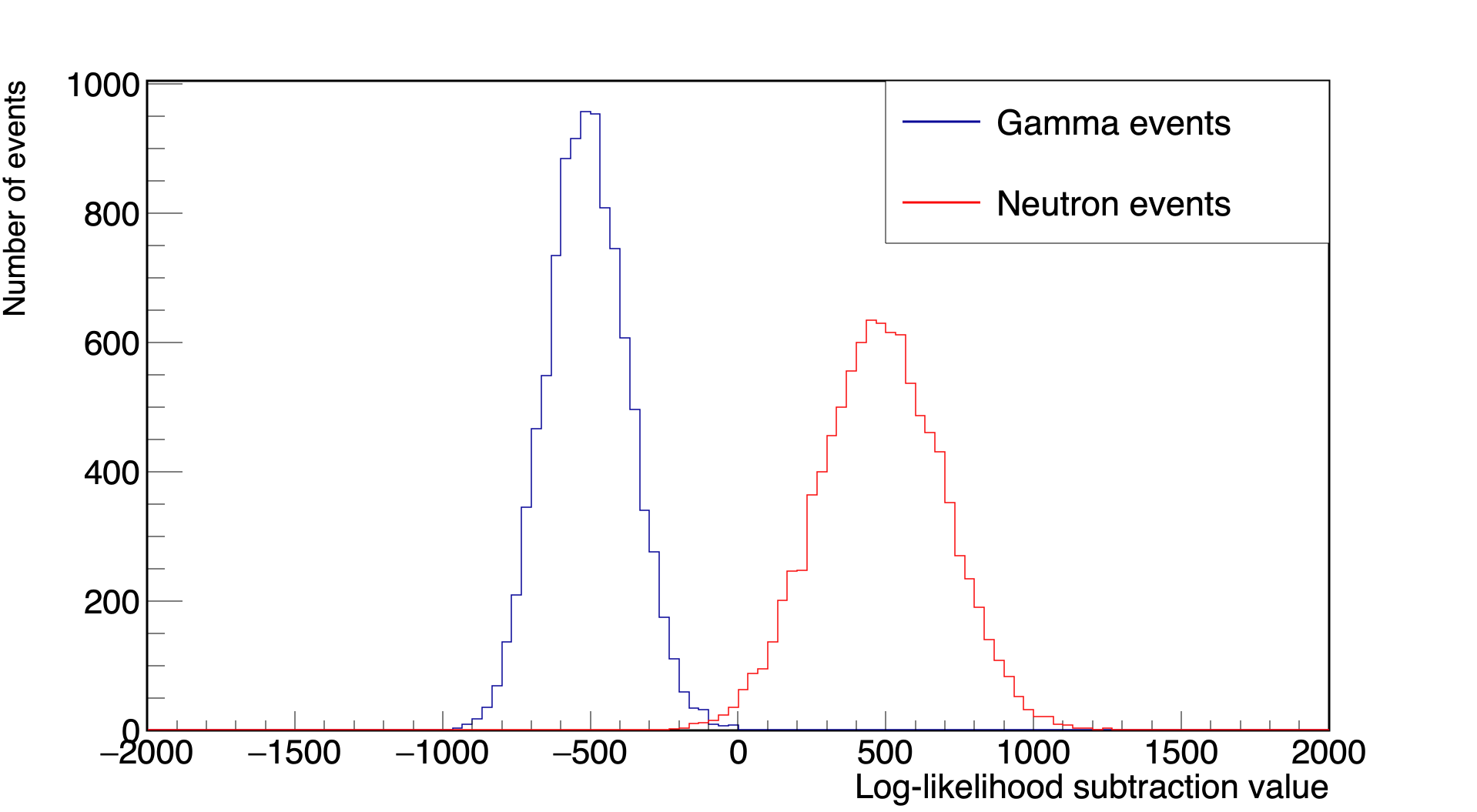}
    }
    
    \caption{Distribution of the Log-likelihood subtraction values. }
    \label{fig:PSD_to_NPE}
\end{figure}

The PSD performance is defined as the ratio of events with a positive likelihood value to the total number of events when the PSD is applied to all events. The accumulated waveforms were converted into PDFs, and the PSD method was applied to each set of simulation data to evaluate the performance of the PSD. Events were analyzed ranging from 10 PE to 1,000 PE, with each containing 10,000 events. The results are presented in Table 2.

%------------------------------------------------------------------------------------------------------------------------------

\section{Conclusion}

\begin{figure}[htbp]
    \centering
    % first subfigure
    \includegraphics[width=0.6\textwidth]{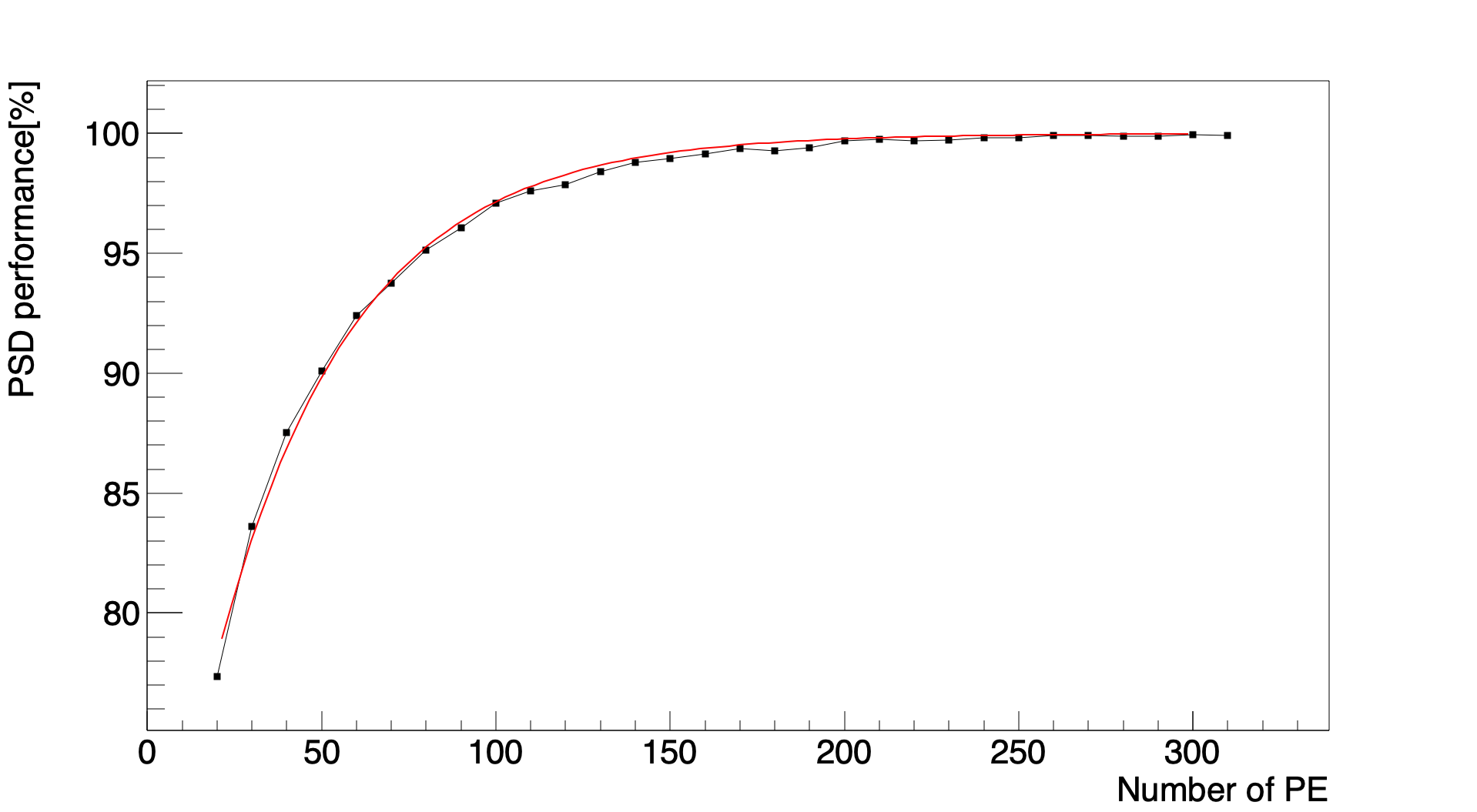}
    \hfill
    \caption{The performance of the PSD according to NPE}
    \label{fig:Performance_of_PSD}
\end{figure}

We determined the minimum NPE required for achieving optimal PSD performance in experiments utilizing 0.1 wt\% Gd-LS with 5 vol\% UltimaGold-F. It was observed that as the number of photoelectrons (NPE) increased, the performance of pulse shape discrimination (PSD) improved. A minimum of 49 photoelectrons (PE) was required to achieve 90\% performance while retaining 97.8\% of gamma-like events.

\section*{Acknowledgment}
%Insert the Acknowledgment text here.
This research was supported by Kyungpook National University Research Fund, 2022.


%\bibliographystyle{ptephy}
%\bibliography{sample}
%
% once the .bbl file has been generated then place the text in your article.

\vspace{0.2cm}
\noindent
%For references,  note how to include DOI information from examples below. 

%This is added by T. Yoneya (editor-in-chief) on 2020/07/09.

\let\doi\relax

%without this code before the command "\begin{thebibliography}{}" , an error will be %flagged. When the bibliography is provided as separate .bib file, then this code %should be placed above the commands "\bibliographystyle{}" and "\bibliography{}" %inside the main TeX file. 

\end{document}